\documentclass[%
twocolumns,
superscriptaddress,
showpacs,preprintnumbers,
 amsmath,amssymb,
 aps,
prc,
longbibliography,
floatfix,
lengthcheck,%
]{revtex4}

\usepackage{amsmath,amssymb}
\usepackage{bm}

\usepackage{graphicx}
\usepackage{dcolumn}
\usepackage{bm}

\newcommand{\beq}{\begin{equation}}
\newcommand{\beqa}{\begin{eqnarray}}
\newcommand{\eeq}{\end{equation}}
\newcommand{\eeqa}{\end{eqnarray}}

\renewcommand{\thanks}{\footnote}

\begin{document}

\preprint{APS/123-QED}

\title{Energy-density-functional calculations including the proton-neutron mixing}

\author{Koichi Sato}
 \affiliation{RIKEN Nishina Center, Wako 351-0198, Japan}

\author{Jacek Dobaczewski}
\affiliation{Institute of Theoretical Physics, Faculty of Physics,University of Warsaw, ul.
Ho{\.z}a 69, 00-681 Warsaw, Poland}
\affiliation{
Department of Physics, PO Box 35 (YFL), FI-40014
University of Jyv{\"a}skyl{\"a}, Finland}

\author{Takashi Nakatsukasa}
\affiliation{RIKEN Nishina Center, Wako 351-0198, Japan}
\affiliation{Center for Computational Sciences, University of Tsukuba,
Tsukuba 305-8571, Japan}

\author{Wojciech Satu{\l}a}
\affiliation{Institute of Theoretical Physics, Faculty of Physics,University of Warsaw, ul.
Ho{\.z}a 69, 00-681 Warsaw, Poland}
\date{\today}

\begin{abstract}
We present results of calculations based on the Skyrme energy density
functional including the arbitrary mixing between protons and
neutrons. In this framework, single-particle states are
superpositions of proton and neutron components and the energy
density functional is fully invariant with respect to
three-dimensional rotations in the isospin space. The isospin of the
system is controlled by means of the isocranking method, which
carries over the standard cranking approach to the isospin space. We
show numerical results of the isocranking calculations performed for
isobaric analogue states in the $A=14$ and $A=40-56$ nuclei. We also
present such results obtained for high-isospin states in $^{48}$Cr,
with constraints on the isospin implemented by using the augmented
Lagrange method.
\end{abstract}

\pacs{21.10.Hw,21.60.Jz,27.40.+z,71.15.Mb}

\maketitle


The superfluidity and superconductivity phenomena manifest themselves
at different physical scales in condensed matter, nuclear, and
elementary particle physics. They reflect the universality of the
underlying Bardeen-Cooper-Schrieffer (BCS) mechanism of forming
Cooper pairs and pair condensates, irrespective of details of attractive
interactions at the Fermi energy.
In atomic nuclei, which are composite systems built of two types
of fermions interacting strongly in charge independent way, the BCS
mechanism offers a unique possibility to form, apart of conventional
neutron-neutron and proton-proton Cooper pairs,  also the
proton-neutron (p-n) pairs of isoscalar ($T=0$) or isovector ($T=1$)
type.

Although the attempts to incorporate the p-n pairing into the
independent-quasi-particle approach date back to the late sixties,
see Refs.~\cite{[Goo79],[Per04w]} and Refs.\ quoted therein for a
review, a consistent theory of nuclear pairing in the vicinity of the
$N=Z$ line, where the p-n correlations are expected to be strongest,
is still missing.  The reason is that the existing approaches
concentrate on introducing the p-n pairing mixing on top of unmixed
proton and neutron single-particle (s.p.) orbitals. However, a
consistent approach to the problem of the p-n pairing requires
implementing the p-n mixing on the mean-field (MF) level, whereby the
s.p.\ wave functions are linear combinations of proton and
neutron components. Basic self-consistency principles require such
mixing to accompany any anticipated p-n mixing on the pairing level.
Moreover, the stability and existence of the p-n pairing condensate
may critically depend on the restoring force related to the p-n
mixing on the MF level, and thus, to obtain meaningful estimates of
the effect, in the theoretical description both must be
simultaneously included.

Our ultimate goal is to develop a consistent symmetry-unrestricted
energy-density-functional (EDF) approach including the p-n mixing both in the
pairing (p-p) and particle-hole (p-h) channels, with rigorous
treatment of the isospin degree of freedom, which is of vital importance
for the understanding of elementary excitations along the $N=Z$ line.
In this work, being a first step in achieving our
goal, we report on a development of the EDF approach
based on extended Skyrme EDF including the p-n mixing
in the p-h channel, with the isospin degree of freedom controlled by means
of three-dimensional isocranking model.

The calculations performed below are based on a local Skyrme EDF
generalized to include the p-n mixing according to the general rules
given by Perli\'nska {\it et al\/}.~\cite{[Per04w]}. At present we
consider only scalar-isoscalar EDFs preserving both the rotational
and isospin symmetries but further generalizations are rather
straightforward. The explicit form of the employed EDF is given in
Eqs. (39), (40), (62), and Table I of Ref.~\cite{[Per04w]}. This
model, hereinafter referred to as pnEDF, was
implemented within the HFODD code (v2.56e)~\cite{[Sch12s]} solving the
nuclear Skyrme-Hartree-Fock(-Bogoliubov) problem by using the
Cartesian deformed harmonic-oscillator basis.

Since the model breaks
distinction between proton and neutron orbitals, the underlying
Kohn-Sham equations must be solved in $A$-dimensional space, where
$A$ denotes the number of nucleon. The neutron and proton numbers $N,Z$
or, alternatively, the third component of the isospin $T_z$, must be
fixed by additional constraints. This is achieved by adding the
isocranking term to the MF Hamiltonian~\cite{[Sat01]}:
\begin{equation}
\hat{h}'=\hat {h}-\vec{\lambda} \cdot \hat{\vec{t}},
\label{isocranking}
\end{equation}
where $\hat {\vec{t}}$ denotes the s.p.\ isospin operator. The
isocranking model is analogous to that used successfully in the
standard tilted-axis-cranking calculations for high-spin states and
can be viewed as the lowest-order approximation to isospin projection
in a sense of Kamlah expansion~\cite{[Kam68]}. By changing the
isocranking frequency $\vec{\lambda}$, we can control the magnitude
and direction of the isospin of the system. In the following, we
shall show  numerical results of the isocranking calculations for
selected isobaric multiplets. In all calculations we shall use the
SkM* parameter set of Ref.~\cite{[Bar82s]}.

For pedagogical reasons, we begin the discussion with the
isocranking calculations without Coulomb interaction, which is the
only source of the isospin-symmetry breaking in our model. Indeed,
at present we disregard the proton-neutron mass difference as
well as any hadronic charge-dependent and charge-symmetry-violating
terms, and thus we limit ourselves to a purely isoscalar EDF. Hence, as long as the
Coulomb energy is switched off, the model Hamiltonian is invariant
under rotations in the isospace and the total and s.p.\ energies become independent
of the direction of the isospin, whereas the s.p.\ wave functions all
acquire common p-n mixing coefficients. This property considerably
simplifies the physics, helps to verify the validity of the code, and, in
particular, helps to work out a strategy of adjusting the
isocranking frequency.

The absolute value (length) of the isocranking frequency is determined from the
results of the standard EDF calculation without the p-n mixing for
non-zero isospin states. The key role is played by the isoaligned states,
$T_z=\pm T$, in the isobaric multiplet. In even-even
nuclei these states correspond to ground-states, which are uniquely
defined by occupying pairwise the Kramers degenerated levels from the
bottom of the potential well and are very well represented by
p-n unmixed Slater determinants~\cite{[Sat09a]}. Hence, in the first
step, we perform standard EDF calculation for the ground state of
$T=T_z$ nucleus and find the difference between the proton and
neutron Fermi energies. The difference sets the length of the
isocranking frequency as $|\vec{\lambda}|=\lambda_z$. With this choice
the resulting proton and neutron Fermi energies roughly coincide, because
the isocranking term, $-\lambda_z \hat{t}_z$, raises the proton
s.p.\ energies by $\lambda_z/2$ and lowers the neutron
energies by $-\lambda_z/2$. Next, by keeping the value of
$\lambda=|\vec{\lambda}|$ fixed, we change gradually the direction of
$\vec{\lambda}$ from $\theta=0^\circ$ to $\theta=180^\circ$ where
$\theta$ denotes the polar angle of $\vec \lambda$,
\begin{equation}
\vec \lambda=(\lambda\sin \theta,0,\lambda\cos\theta). \label{eq:shiftedsemicircle2}
\end{equation}
In this way we
gradually reduce $T_z$ or, alternatively, decrease $N$ and increase
$Z$. The results are entirely independent of the azimuthal angle
$\phi$ in the isospace.

\begin{figure}[htb]
\begin{center}
\includegraphics[width=0.45\textwidth]{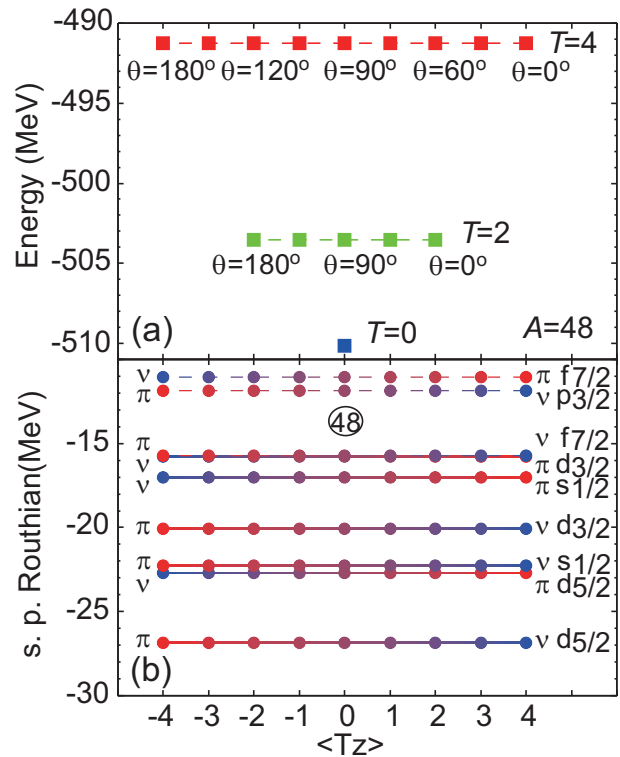}
\end{center}
\caption{
(Color online) (a) Energies of the $T = 0$, 2, and 4 states in $A=48$
isobars obtained by the isocranking calculation without the Coulomb
interaction as functions of $\langle \hat T_z\rangle$. (b) The
s.p.\ Routhians for $|\vec{\lambda}|=11$ MeV ($T=4$ states)
calculated without the Coulomb interaction as functions of $\langle
\hat T_z\rangle$. There is no p-n mixing at $T_z=4$ and -4, where
each s.p.\ orbital is either proton ($\pi$)
or neutron ($\nu$) state. Note that the $\pi$-$\nu$ character at $T_z=-4$
is opposite to that at $T_z=4$.
}
\label{fig:A=48noCoulomb}
\end{figure}

The procedure is illustrated for a representative example of $A=48$
nuclei. In Fig.~\ref{fig:A=48noCoulomb}(a), we plot the total
energies of the $T=2$ and $T=4$ states in $A=48$ nuclei obtained within
the isocranking calculations together with that of the $T=0$ state
(the ground state of $^{48}$Cr). For  the $T=2$ and $T=4$ states, we
take the standard EDF solutions of $^{48}$Ti and $^{48}$Ca,
respectively, as initial states and iteratively solve the isocranking
pnEDF equations. The direction of isospin is controlled by changing
the direction of $\vec{\lambda}$ while keeping its magnitude
$|\vec{\lambda}|$ fixed and equal to 11\,MeV (6\,MeV) for $T=4$
$(T=2)$, respectively. Let us recall that these values are inferred
from the differences in proton and neutron Fermi energies obtained in
the p-n unmixed EDF calculations in $^{48}$Ca and $^{48}$Ti,
respectively. Going from $^{48}$Ca ($T_z=4$) to $^{48}$Ni ($T_z=-4$)
or from $^{48}$Ti ($T_z=2$) to $^{48}$Fe ($T_z=-2$), $\vec \lambda$
draws a semicircle trajectory ($\theta=0^\circ \rightarrow
180^\circ$) whose radius is $|\vec \lambda|$ with the center at the
origin. Fig.~\ref{fig:A=48noCoulomb}(a) shows the results for
different values of the polar angles of isocranking frequency
$\theta$. No p-n mixing takes place in the states with $T=|T_z|$,
such as $^{48}$Ca ($T=T_z=4$), $^{48}$Ti ($T=T_z=2$) $^{48}$Ni
($T=-T_z=4$), $^{48}$Fe ($T=T_z=2$), and $^{48}$Cr ($T=T_z=0$). These
correspond to either $\theta=0^\circ$ or $180^\circ$.

In these calculations, the azimuthal angle $\phi$ was set to
$0^\circ$. We have confirmed that the results do not depend on $\phi$
even if the Coulomb interaction is included. This is so, because the
Coulomb interaction is aligned along the $T_z$ axis and thus commutes
with isorotations around this axis. Therefore, in what follows the
isocranking is investigated only in the $T_z-T_x$
($\lambda_z-\lambda_x$) plane. This choice has an additional
advantage of avoiding the time-reversal symmetry breaking inherent to
isorotations along the $T_y$ axis, see discussion in
Ref.~\cite{[Per04w]}.

The procedure outlined above can also be used to calculate odd-$T$
states, see also Fig.~1 in Ref.~\cite{[Sat01a]}. In $A=4n$ cases it
would require to carry out first the standard EDF calculation for an
odd-odd nucleus with $T=T_z$, e.g., in $^{48}_{21}$Sc$_{27}$ for the
$A=48$ isobars discussed above. Having the initial state, the odd-$T$
multiplet can be calculated by following essentially the same
isocranking procedure as used for the even-$T$ cases. However, the
calculation of odd (odd-odd) nuclei involves breaking of the
time-reversal symmetry. In order to avoid complications caused by the
time-reversal symmetry non-conservation which, among the others,
creates ambiguities in configuration assignment of the initial state
we shall concentrate here on the isobaric multiplets obtained by
applying isocranking to the ground states of even-even  $T=|T_z|$
nuclei that preserve the time-reversal symmetry. In $A=4n+2$ nuclei,
this assumption will limit our calculations to odd-$T$ isobaric
multiplets. In particular, in odd-odd $N=Z$ nuclei, it will allow us to
represent the $T=1$ IASs by means of a single time-symmetry-conserving
(anti-aligned in space) Slater determinant. More general situations
will be discussed elsewhere.

In Fig.~\ref{fig:A=48noCoulomb}(a) we plot the IASs obtained by
isocranking the $T=T_z$ states. As expected from the isospin symmetry,
the energy curves are flat, that is, independent of the direction of the
isospin or value of $T_z$. We have independently verified that
the non-zero isospin states shown in Fig.~\ref{fig:A=48noCoulomb}(a)
can be obtained by isocranking the initial $T=0$
state, that is, by starting with the $T=0$ state and performing
the $T_x$-axis isocranking. Then, the term $-\lambda_x \hat{t}_x$ gives {\it
vertical\/} excitations: $T=0\rightarrow T=2\rightarrow T=4$ with
$\langle \hat T_z \rangle=0$. Similarly, the $T_z$-axis isocranking term, $-\lambda_z
\hat{t}_z$, applied to the same $T=0$ state produces a sequence of
{\it diagonal\/} excitations: $T=0\rightarrow T=2\rightarrow T=4$
with $T=|\langle \hat T_z \rangle|$. Irrespective of the direction
of the isocranking axis, one always obtains only even-$T$ states. This
is because, owing to the time-reversal symmetry, with increasing
isocranking frequency only a configuration change with $\Delta T=2$
occurs at each level crossing. To obtain odd-$T$ states, explicit 1p-1h
excitations are required~\cite{[Sat01]}.

Figure \ref{fig:A=48noCoulomb}(b) shows s.p.\ Routhians,
that is, the eigenenergies of $\hat h^\prime$ (\ref{isocranking}),
calculated for $T=4$ states as functions of $\langle \hat
T_z\rangle$. We can clearly see that they are
independent of $\langle \hat T_z \rangle$. The Routhians are pure
proton or neutron s.p.\ states only at $\theta= 0^\circ$ and
$180^\circ$. At all other tilting angles $0^\circ < \theta
<180^\circ$ the Routhians are p-n mixed.

Quantitative features of the p-n mixing are listed in
Table~\ref{tab:no_Coulomb} for the two highest occupied s.p.\
spherical orbitals. In the discussion it is enough to focus on one
state, say, the $f_{7/2}$ state, which, at $\theta=0^\circ$,
corresponds to a pure neutron state. Gradual tilting of the
isorotation axis increases proton component of the state, which, at
$\theta=90^\circ$, reaches exactly 50\%. Eventually, at
$\theta=180^\circ$ the state becomes a pure proton state. This result
applies to all orbitals, which interchange their character
from pure proton/neutron at $\theta=0^\circ$ to pure neutron/proton
at $\theta=180^\circ$ and are fifty-fifty mixed exactly at
$\theta=90^\circ$. It is worth stressing that when the Coulomb
interaction is switched off, the total isospin is exactly parallel to
$\vec \lambda$. Moreover, the isospin of each s.p.\ state is either
parallel or anti-parallel to $\vec{\lambda}$. The s.p.\ eigenstates
of Eq.~(\ref{isocranking}) are, at the same time, the eigenstates of
$\hat{h}$, whose eigenvalues are also independent of $\langle \hat
T_z \rangle$. As we shall see later, these features occur only when
the Coulomb interaction is neglected.

\begin{table}[b]
\caption{\label{tab:table1}
Calculated expectation values of twice the isospin operators
$2\hat{t}_x$ and $2\hat{t}_z$ for the two highest occupied orbitals of
the  $T=4$ states in $A=48$ isobars. The Coulomb interaction is
switched off. The expectation values of the $T_x$ and $T_z$ components of
the total isospin as well as its polar angles $\theta_T$ are also shown.
}
\label{tab:no_Coulomb}
\begin{ruledtabular}
\begin{tabular}{crrrrrrrrrr}
$\theta$&	& $0^\circ$    	& $60^\circ$   	& $90^\circ$   	& $120^\circ$  	& $180^\circ$ \\
$\theta_T$&     & $0^\circ$     & $60^\circ$    & $90^\circ$    & $120^\circ$   & $180^\circ$ \\
\colrule
$f_{7/2}$     &  $\langle \hat \tau_x \rangle $  & 0.00 & 0.87 & 1.00 &  0.87 & 0.00 \\
              &  $\langle \hat \tau_z \rangle $  & 1.00 & 0.50 & 0.00 & -0.50 &-1.00 \\
$d_{3/2}$     &	 $\langle \hat \tau_x \rangle $  & 0.00 &-0.87 &-1.00 &	-0.87 &	0.00 \\
              &  $\langle \hat\tau_z \rangle  $  &-1.00 &-0.50 & 0.00 &  0.50 &	1.00 \\
\colrule
$\langle \hat T_x \rangle$ 	    &            & 0.00 & 3.46 & 4.00 &  3.46 & 0.00 \\
$\langle \hat T_z \rangle$ 	    &            & 4.00 & 2.00 & 0.00 & -2.00 &-4.00 \\
\end{tabular}
\end{ruledtabular}
\end{table}

\begin{figure}[htb]
\begin{center}
\includegraphics[width=0.45\textwidth]{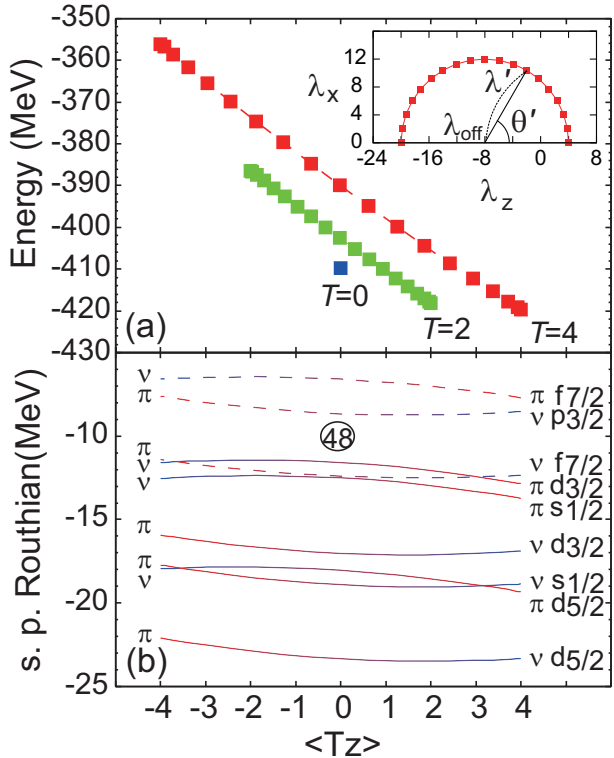}
\end{center}
\caption{(Color online)
(a) Energies of the $T \simeq 0$, 2, and 4 states in $A=48$ isobars
obtained by the isocranking calculation with the Coulomb interaction
included. Results for every 10$^{\circ}$ of $\theta^\prime$ are
plotted. The inset shows the isocranking frequencies used for the
$T\simeq 4$ calculation in the $\lambda_x-\lambda_z$ plane. (b) Same
as Fig.~\protect\ref{fig:A=48noCoulomb}(b) but with the Coulomb
interaction included.
}
\label{fig:A=48Coulomb}
\end{figure}

\begin{table}[tb]
\caption{\label{tab:table2}
Same as Table I but calculated with  the Coulomb interaction.
Note that the polar angles $\theta$ of $\vec\lambda$,
defined in Eq.~(\protect\ref{eq:shiftedsemicircle2})
are very different from the polar angles $\theta^\prime$ of $\vec\lambda^\prime$,
defined in Eq.~(\protect\ref{eq:shiftedsemicircle}), and that the latter are close
to the polar angles $\theta_T$ of $\langle \hat{\vec T} \rangle$.
}
\label{tab:with_Coulomb}
\begin{ruledtabular}
\begin{tabular}{crrrrrrrrrr}
 $\theta       $&                                               & $0^\circ$   & $100.9^\circ$ & $108.4^\circ$ & $133.9^\circ$ & $180^\circ$   \\
 $\theta^\prime$&	                                        & $0^\circ$   & $60^\circ$   	& $90^\circ$   	& $120^\circ$  	& $180^\circ$   \\
$\theta_T$&      & $0^\circ$ &  $62.3^\circ$ & $90.3^\circ$   &   $118.0^\circ$ &      $180^\circ$  \\
\colrule
$f_{7/2}$     & $\langle \hat \tau_x \rangle $  &	 0.00	&	 0.89	&	1.00	&	 0.88	&       0.00	\\
              &  $\langle \hat \tau_z \rangle $ &	 1.00	&	 0.46	&	0.01	&	-0.47	&      -1.00	\\
$d_{3/2}$     & $\langle \hat \tau_x \rangle $ 	&	 0.00	&       -0.90	&       -1.00	&	-0.88	&       0.00	\\
              & $\langle \hat\tau_z \rangle $ 	&	-1.00	&       -0.44	&	0.00	&	 0.47	&       1.00	\\
\colrule
$\langle \hat T_x \rangle$ 	    &           &       0.00	&	 3.54	&	4.00	&	 3.53	&       0.00	\\
$\langle \hat T_z \rangle$ 	    &        	&	4.00	&	 1.86	&	-0.02	&	-1.88	&      -4.00	\\
\end{tabular}
\end{ruledtabular}
 \end{table}

Next, we discuss calculations performed with the Coulomb interaction
included and treated exactly both in the direct and exchange
channels~\cite{[Dob09ds]}. To reduce the Coulomb repulsion, the
Coulomb interaction always tends to increase (decrease) the neutron
(proton) components of all states; therefore states with larger
values of $\langle T_z \rangle$ are always energetically favored.
Moreover, the s.p.\ Routhians now vary as functions of the tilting
angle $\theta$, and in the interval of $0^\circ \leq \theta \leq
180^{\circ}$ many level crossings may take place. This often becomes
a source of the {\it ping-pong\/} divergence during the iteration
\cite{[Dob00c]}, rising technical difficulties in the applications.

A simple and efficient prescription to avoid the level
crossings of Routhians is to parametrize the isocranking
frequencies as
\begin{equation}
\vec \lambda=(\lambda^{\prime}\sin \theta^\prime,0,\lambda^{\prime}\cos\theta^\prime+\lambda_{\rm off}). \label{eq:shiftedsemicircle}
\end{equation}
The semicircle trajectory, which we introduced above, is now shifted
in the $T_z$-direction of the isospace by the offset value of $\lambda_{\rm
off}$, as depicted in the inset of Fig.~\ref{fig:A=48Coulomb}(a). The
rationale for this choice is in the fact that the Coulomb interaction
depends on the $T_z$-component of the isospin. It can thus be
regarded as an effective additional contribution to the isocranking
term in the $T_z$-direction. The offset $\lambda_{\rm off}$ in
Eq.~(\ref{eq:shiftedsemicircle}) is added to compensate for this
additional contribution.

In practice, we proceed as follows: First,
we perform the standard EDF calculation for the $T\simeq |T_z|$ nuclei,
so as to specify the endpoints $\lambda_{\rm off} \pm
\lambda^{\prime}$ at $\theta^\prime = 0^\circ$ and $180^{\circ}$.
Again, they are determined from the differences of the proton and
neutron Fermi energies, as was the case without the Coulomb interaction.
For example, in the representative example of the $T \simeq 4$ states
in $A=48$ isobars, in $^{48}$Ca we obtain $\lambda_n-\lambda_p \simeq
4$\,MeV, and in $^{48}$Ni we obtain $\lambda_n-\lambda_p \simeq
-20$\,MeV. Hence, we take $\lambda_{\rm off}=-8$ and
$\lambda^\prime=12$\,MeV . The same procedure applied to $T\simeq 2$
multiplet gives $(\lambda_{\rm off},\lambda^\prime)=(-8.0,6.4)$\,MeV.

Figure \ref{fig:A=48Coulomb}(a) shows calculated energies of
the $T \simeq 0,2$, and 4 multiplets in $A=48$ isobars. Approximate
linear dependence of the total energies on $\langle \hat T_z \rangle$
results from a dominance of the isovector part of the Coulomb
interaction, over its isotensor component, in the isospin
symmetry-breaking mechanism. In terms of the Coulomb energy, which
macroscopically behaves as $Z^2=T_z^2-AT_z+A^2/4$, this fact reflects a
dominant role of the linear term $-AT_z$. It is worth stressing that,
in the presence of the isospin-symmetry breaking fields, the rotation in
the isospace becomes nonuniform. Indeed, as shown in
Table~\ref{tab:with_Coulomb}, the direction of isospin is no longer
parallel to $\vec\lambda$ (angle $\theta$). However, the direction
of the shifted cranking axis $\vec\lambda'$ (angle $\theta'$) is
approximately parallel with the direction of the isospin meaning that
the use of shifted semicircle can be viewed as effective way of
restoring the isospin symmetry in the vicinity of the Fermi energy.

\begin{figure}[htb]
\begin{center}
\includegraphics[width=\columnwidth]{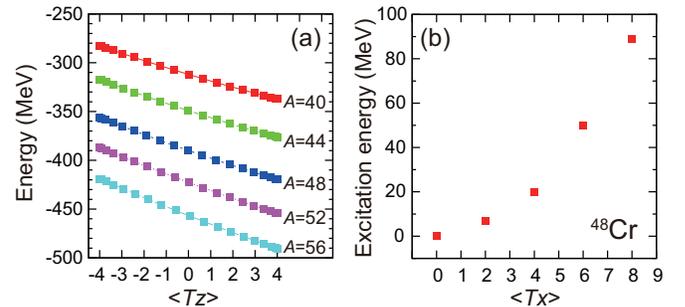}
\end{center}
\caption{(Color online)
Left, energies of the $T \simeq 4$ states in $A=40-56$ isobars with even
$A/2$ as functions of $\langle \hat T_z\rangle$. For $A=40$, 44, 48,
52, and 56, we used the values of $(\lambda_{\rm off} ,\lambda^\prime
)=(-6.8,13.6)$, $(-7.4, 12.4)$, $(-8.0, 12.0)$, $(-8.4, 11.6)$, and
$(-8.8,12.4)$\,MeV, respectively. Right,
excitation energies of $^{48}$Cr for $\langle \hat
T_x\rangle=0$, 2, 4, 6, and 8, calculated by using the augmented Lagrange
method.
}
\label{fig:a040-056}
 \end{figure}

The usefulness of the shifted semicircle technique is illustrated in
Fig.~\ref{fig:A=48Coulomb}(b) and Fig.~\ref{fig:a040-056}(a).
Fig.~\ref{fig:A=48Coulomb}(b) shows the s.p.\ Routhians in function
of $\langle \hat T_z\rangle$ calculated for the $T \simeq 4$ states
in $A=48$ isobars. In this case, the endpoint nuclei
$^{48}_{28}$Ca$_{20}$ and $^{48}_{20}$Ni$_{28}$ are doubly-magic. The
shifted semicircle method causes the neutron and proton Fermi
energies and, in turn, the proton and neutron magic gaps, to match one
another, independently, to a large extent, of values of $\langle \hat
T_z\rangle$. As a result, for all values of $\langle \hat
T_z\rangle$, a sizable gap at $A=48$ stays open in the s.p.\ spectrum
and the crossings are avoided. Fig.~\ref{fig:a040-056}(a) shows the
results obtained for the $T \simeq 4$ multiplets in $A=40-56$
isobars. Systems with $|T_z|=4$ and $A \neq 48$ are not magic nuclei,
so there is no large shell gap at the Fermi surface in their s.p.\
spectra. Nevertheless, the shifted semicircle method works nicely.

The isocranking calculations are based on a simple linear constraint
method, whereupon fixed values of the isocranking frequencies
$\vec\lambda$ or $\vec\lambda'$ lead, in general, to non-integer
values of $\langle\hat{T}_z\rangle$. To improve on that, we also
implemented in our code a method for optimization with constraints,
known as the augmented Lagrange method~\cite{[Sta10],[Sch12s]}, which
is  widely used in quantum chemistry. Using this method, one can
obtain states with $\langle \hat{T}_z\rangle$ having exactly integer
values. In Fig.~\ref{fig:a040-056}(b), we plot the excitation energies
of the $T \simeq 0$, 2, 4, 6, and 8 states of $^{48}$Cr, calculated
with the augmented Lagrange method. In these calculations, we employed
the constraints on $\langle \hat{T}_z\rangle=0$ and $\langle\hat{T}_x
\rangle = 0$, 2, 4, 6 ,and 8, respectively. The obtained
(quasi)parabolic behavior of the calculated energies is determined
mostly by the nuclear symmetry energy. One should bear in mind,
however, that the energies along the parabola also depend on shell
effects, manifesting themselves, among the other, in rapid shape
changes. The calculated quadrupole deformations for states with
$\langle\hat{T}_x \rangle =0$, 2, 4, 6, and 8 are equal to
$\beta_2=0.27$, 0.19, 0, 0.07, and 0, respectively. The latter four
values are in perfect agreement with the ground state equilibrium
deformations calculated directly by using the EDF method in the
$^{48}$Ti, $^{48}$Ca, $^{48}$Ar, and $^{48}$S nuclei, respectively,
which are their $T\simeq T_z$ IASs. This constitutes an explicit
illustration of the fact that the isospin symmetry and isospin quantum number,
albeit approximate, are very powerful concepts in nuclear structure.

To conclude, we consider the case of $A=4n+2$ nuclei, focusing our attention
on the well-known $I=0^+, T=1$ triplet of states in the $A=14$ isobars~\cite{[Boh69]}.
Here, two members of the triplet, namely the $I=0^+, T \simeq |T_z|=1$
ground states of $^{14}$C and $^{14}$O nuclei are represented in our model by
the standard EDF ground states without the p-n mixing.
Their $T_z=0$ IAS representing the excited $I=0^+,T=1$ state in
$^{14}$N, is calculated by using the isocranking model, and is described by a
single time-even Slater determinant built of s.p.\ p-n mixed orbitals.
Let us recall that in the case of $A=4n+2$ nuclei
the odd-$T$ states are represented as the time-even Slater determinants, whereas
the even-$T$ states break the time-reversal symmetry.

\begin{figure}[htb]
\begin{center}
\includegraphics[width=0.7\columnwidth]{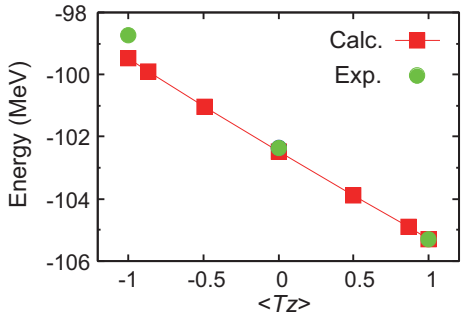}
\end{center}
\caption{(Color online)
Energies of $T \simeq 1$ states in $A=14$ isobars in comparison with
the experimental data \cite{[Brookhaven]}. The calculated energies
are shifted by 3.2 MeV such that the experimental and calculated
binding energies of $^{14}$C coincide. The results of isocranking
calculations for every $30^\circ$ of $\theta^\prime$ between
$\theta^\prime =0^\circ$ and $180^\circ$ are plotted.
}
\label{fig:c014}
 \end{figure}
The calculated energies and the corresponding experimental data are
depicted in Fig.~\ref{fig:c014}. To correct for a deficiency of the SkM*
EDF, which overestimates the binding energy of $^{14}$C, the entire theoretical curve is
shifted up by 3.2\,MeV. In $^{14}$N, the calculated state,
representing the excited $0^+,T=1$ state (the IAS with
$T_z=0$), is described by a single Slater determinant built of the p-n mixed
s.p.\ states. The isocranking calculation correctly reproduce its energy relative
to the IAS with $T_z=1$. This proves that the model
is indeed capable of quantitatively describing the excitation energies
of the $0^+, T=1$ IASs. This can be contrasted with single-reference p-n unmixed EDF
models, wherein such states do not exist at all~\cite{[Sat10]}.

In summary, for the first time we solved the generalized Skyrme Kohn-Sham equations
including the arbitrary mixing of protons and neutrons. Values of the
average total isospin and its $T_x$, $T_y$, and $T_z$ components were
controlled by using the isocranking approximation. The model was
applied to the isobaric analogue states in even-even $A=40-56$ and
odd-odd $A=14$ isobars. We have demonstrated that the p-n mixed
single-reference EDF approach is capable of quantitatively describing
the isobaric analogue excited states. Of particular importance are
the $I=0^+, T=1, T_z =0$ states in odd-odd $N=Z$ nuclei because of
their participation in the superallowed Fermi beta decay that is used
to test the Standard Model of particle physics~\cite{[Tow10]}. We
have also shown the results of the augmented Lagrange method for
high-isospin states in $^{48}$Cr. Such calculations can be used to
study the nuclear symmetry energy.

This work is a first step toward the EDF calculations including the
p-n mixing in both p-p and p-h channels. Such simultaneous mixing in
both channels is required by fundamental self-consistency arguments.
When implemented, it will allow us to reexamine, from a completely
new perspective, the fundamental questions concerning formation and
stability of static p-n pairing correlations, coexistence of
different condensates, microscopic nature of the Wigner energy and
properties of the symmetry energy in paired systems.

As discussed in Ref.~\cite{[Sat10]}, the MF approach is affected by
spurious isospin mixing, which can be quantified and removed by
performing the isospin projection and subsequent Coulomb
rediagonalization. The implementation of the isospin projection of
pnEDF states is now in progress. Furthermore, by using the technology
of iterative methods
\cite{[Nak07a],[Ina09x],[Toi10s],[Avo11],[Sto11s],[Car12a],[Hin13]},
we plan to construct an efficient numerical code solving equations of
the quasiparticle random phase approximation based on the p-n mixed
EDF. This may open new possibilities of investigating collective
excitations of the isobaric analogue states and charge-exchange
reactions to/from these states.

This work was partly supported by JSPS KAKENHI (Grant numbers
20105003, 21340073, and 25287065), by the Academy of Finland and
University of Jyv\"askyl\"a within the FIDIPRO programme, and by the
Polish National Science Center under Contract No.\ 2012/07/B/ST2/03907.
The numerical calculations were
carried out on SR16000 at Yukawa Institute for Theoretical Physics in
Kyoto University and on RIKEN Integrated Cluster of Clusters (RICC)
facility.  We also acknowledge the CSC - IT Center for Science Ltd,
Finland, for the allocation of computational resources.

\bibliographystyle{unsrt}

\end{document}